\documentclass[longauth]{aa} 
\usepackage{graphicx}
\usepackage{graphics}
\usepackage{colortbl}
\usepackage{txfonts}
\usepackage{natbib}
\begin{document}
  \title{Energy dependent $\gamma$-ray morphology in the Pulsar wind
  nebula HESS\,J1825--137}
  \author{F. Aharonian\inst{1}
    \and A.G.~Akhperjanian \inst{2}
    \and A.R.~Bazer-Bachi \inst{3}
    \and M.~Beilicke \inst{4}
    \and W.~Benbow \inst{1}
    \and D.~Berge \inst{1}
    \and K.~Bernl\"ohr \inst{1,5}
    \and C.~Boisson \inst{6}
    \and O.~Bolz \inst{1}
    \and V.~Borrel \inst{3}
% ...
  }
  \offprints{Stefan Funk (Stefan.Funk@mpi-hd.mpg.de)}

  \institute{
    Max-Planck-Institut f\"ur Kernphysik, P.O. Box 103980, D 69029
    Heidelberg, Germany
    \and
    Yerevan Physics Institute, 2 Alikhanian Brothers St., 375036 Yerevan,
    Armenia
    \and
    Centre d'Etude Spatiale des Rayonnements, CNRS/UPS, 9 av. du Colonel Roche, BP
    4346, F-31029 Toulouse Cedex 4, France
    \and
    Universit\"at Hamburg, Institut f\"ur Experimentalphysik, Luruper Chaussee
    149, D 22761 Hamburg, Germany
    \and
    Institut f\"ur Physik, Humboldt-Universit\"at zu Berlin, Newtonstr. 15,
    D 12489 Berlin, Germany
    \and
    LUTH, UMR 8102 du CNRS, Observatoire de Paris, Section de Meudon, F-92195 Meudon Cedex,
    France
% ...
}


  \abstract {}
      {We present results from deep $\gamma$-ray observations of the
	Galactic pulsar wind nebula \object{HESS\,J1825--137}\
	performed with the H.E.S.S.\ array.}
      { Detailed morphological and spatially resolved spectral
	studies reveal the very high-energy (VHE) $\gamma$-ray aspects
	of this object with unprecedented precision. }
      { We confirm previous results obtained in a survey of the
	Galactic Plane in 2004. The $\gamma$-ray emission extends
	asymmetrically to the south and south-west of the energetic
	pulsar \object{PSR\,J1826--1334}, that is thought to power the
	pulsar wind nebula. The differential $\gamma$-ray spectrum of
	the whole emission region is measured over more than two
	orders of magnitude, from 270~GeV to 35~TeV, and shows
	indications for a deviation from a pure power law. Spectra
	have also been determined for spatially separated regions of
	HESS\,J1825--137. The photon indices from a power-law fit in
	the different regions show a softening of the spectrum with
	increasing distance from the pulsar and therefore an energy
	dependent morphology.}
      { This is the first time that an energy dependent morphology has
	been detected in the VHE $\gamma$-ray regime.
	The VHE $\gamma$-ray emission of HESS\,J1825--137 is
	phenomenologically discussed in the scenario where the
	$\gamma$-rays are produced by VHE electrons via Inverse
	Compton scattering. The high $\gamma$-ray
	luminosity of the source cannot be explained on the basis
	of constant spin-down power of the pulsar and requires
	higher injection power in past.}
  \authorrunning{F. Aharonian et al.}  
  \titlerunning{Energy dependent $\gamma$-ray morphology in
      HESS\,J1825--137}

  \maketitle


\section{Introduction}
A growing number of extended objects that seem to be associated with
energetic pulsars are detected in the Galactic Plane by their very
high-energy (VHE, energy $E_{\gamma} \gtrsim 100$~GeV) $\gamma$-ray
emission. Latest results on this class of objects include emission
from \object{MSH--15--5\emph{2}}
(\object{HESS\,J1514--591})~\citep{HESSMSH} and \object{Vela~X}
(\object{HESS\,J0835--455})~\citep{HESSVelaX}, and the two sources in
the \object{Kookaburra} region (\object{HESS\,J1420--607} and
\object{HESS\,J1418--609}) as described in~\citet{HESSKookaburra}.  If
these associations are correct, then these objects are pulsar wind
nebulae (PWN), objects generally thought to be powered by a
relativistic particle outflow (electrons and positrons) from a central
source. The central source -- a pulsar -- is a rapidly rotating
neutron star generated in a supernova event. The relativistic
wind of particles flows freely out until its pressure is
balanced by that of the surrounding medium. In that region the
wind decelerates and a standing termination shock is formed at which
particles are accelerated~\citep{KennelCoroniti, AhaAtoKif97}. The
existence of electrons accelerated to energies $>100$~TeV in such PWN
has been established by X-ray observations of synchrotron emission,
e.g. in the Crab nebula~\citep{ChandraCrab}. VHE $\gamma$-rays can be
generated in PWN from the high-energy electrons by non-thermal
bremsstrahlung or inverse Compton (IC) scattering on photon target
fields, such as the cosmic microwave background (CMBR) or star-light
photons.
\begin{figure*}
  \centering
  \includegraphics[width=0.7\textwidth]{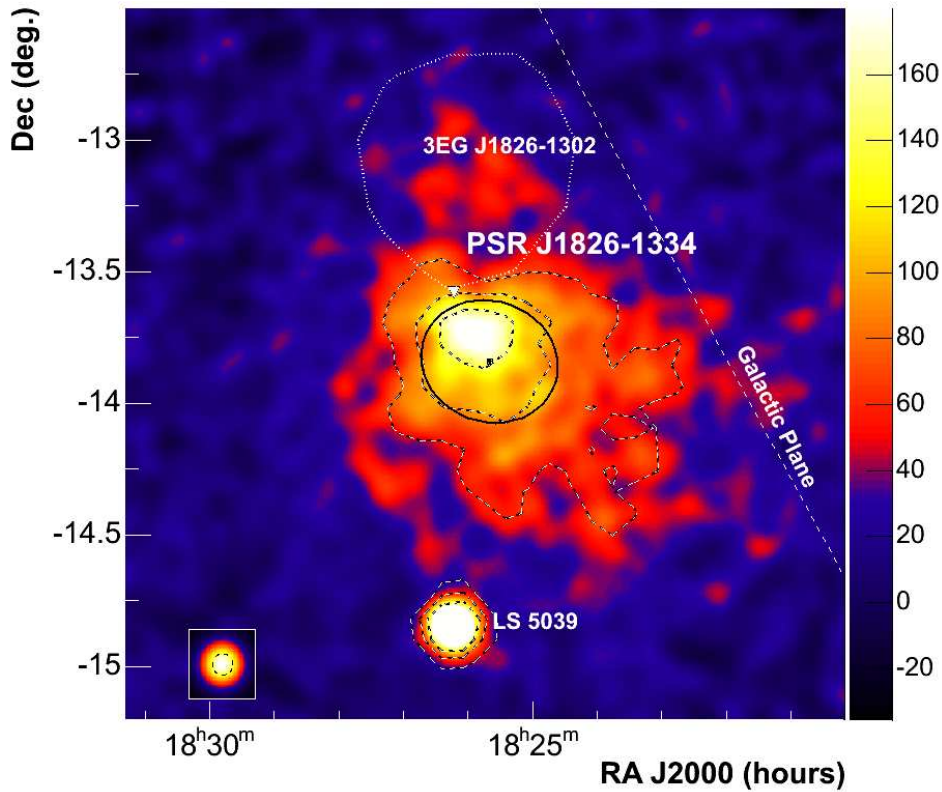}
  \caption{Acceptance-corrected smoothed excess map (smoothing radius
  2.5\arcmin) of the $2.7\degr\ \times 2.7\degr$ field of view
  surrounding HESS\,J1825--137. The linear colour scale is in units of
  integrated excess counts within the smoothing radius of
  2.5\arcmin. The excess has been derived from a model of the system
  acceptance as described in the text. The inset in the bottom left
  corner shows the PSF of the dataset (smoothed in the same way as the
  excess map with the black dashed circle denoting the smoothing
  radius). The dashed black and white contours are linearly spaced and
  denote the $5 \sigma$, $10 \sigma$ and $15 \sigma$ significance
  levels (the $5 \sigma$ contour being the outermost one), determined
  with a $\theta = 0.1\degr$ radius cut. The best fit position of
  HESS\,J1825--137 is marked with a black square, the best extension
  and position angle by a black ellipse (see text). The dotted white
  contour shows the 95\% positional confidence contour of the
  unidentified EGRET source \object{3EG\,J1826--1302}. The
  position of the pulsar PSR\,J1826--1334 is marked by a white
  triangle. The bright point-source to the south of HESS\,J1825--137
  is the microquasar \object{LS\,5039}\
  (\object{HESS\,J1826--148})~\citep{HESS5039}. The colour scale for
  this source is truncated in this Figure. The Galactic plane is shown
  as a white dashed line. Some indication for an additional emission
  region to the north of the pulsar is seen.}
  \label{fig::skymap}
\end{figure*}

One such object, HESS\,J1825--137, has been detected by the High
Energy Stereoscopic System (H.E.S.S.) in a survey of the inner
Galaxy~\citep{HESSScan, HESSScanII} and has subsequently been
associated with the X-ray PWN G18.0--0.7 surrounding the energetic
pulsar PSR\,J1826--1334~\citep{HESS1825}. This pulsar PSR\,J1826--1334
(also known as PSR\,B1823--13) was detected in the Jodrell Bank 20~cm
radio survey~\citep{RadioDetection} and is among the 20 most energetic
pulsars in the current ATNF catalogue (spin down power ${\dot E} = 3
\times 10^{36}$ erg/s). The distance of PSR\,J1826--1334
as measured from the dispersion of the radio pulses is
$3.9\pm0.4$~kpc~\citep{Cordes_Lazio}. The radio detection further
revealed characteristic properties of the system that are similar to
those of the well studied Vela pulsar, namely a pulse period of
101\,ms and a characteristic age of 21.4~kyears (derived by $\tau =
P/2 \dot{P}$). This age renders PSR\,J1826--1334 one of the 40
youngest pulsars detected so far~\citep{ATNF}, and due to this, deep
radio observations were performed to find emission associated with the
remnant of the Supernova explosion that gave rise to the
pulsar. However, deep VLA observations of the 20\arcmin\, surrounding
the pulsar have failed to detect this Supernova remnant
(SNR)~\citep{VLABraun}.

Initial observations of the region in X-rays with
ROSAT~\citep{ROSAT1825} revealed a point source surrounded by an
elongated diffuse region of size $\sim$5\arcmin. The X-ray emission
region was subsequentially observed with the ASCA instrument and the
data confirmed the picture of a compact object surrounded by an
extended emission region~\citep{ASCA1825}. While ROSAT data did not
provide sufficient statistics, ASCA data lacked the spatial resolution
to resolve and interpret the sources in this region. The situation was
clarified in an XMM-Newton observation in which high angular
resolution observations revealed a compact core of extension
30\arcsec\, surrounding PSR\,J1826--1334, and furthermore an
asymmetric diffuse nebula extending at least 5\arcmin\, to the south
of the pulsar~\citep{XMM1825}. In this XMM-Newton dataset the signal
to noise ratio deteriorates rapidly at offsets larger than 5\arcmin\,
and for this reason the XMM data cannot place useful constraints on
the presence of a faint shell of emission at larger radii as might be
produced by an associated SNR. The extended asymmetric structure was
attributed to synchrotron emission from the PWN of
PSR\,J1826--1334~\citep{XMM1825}. The X-ray spectrum in the diffuse
emission region follows a power law with photon index $\Gamma \sim$2.3
and an X-ray luminosity between 0.5 and 10~keV of $L_x \sim3 \times
10^{33}$ erg s$^{-1}$ compared to the X-ray spectrum for the compact
core following a power law with $\Gamma \sim$1.6 and $L_x \sim9 \times
10^{32}$ erg s$^{-1}$ (these luminosities are derived assuming a
distance of 4~kpc). \citet{XMM1825} discussed various scenarios to
explain the asymmetry and offset morphology of the PWN G18.0--0.7. The
most likely explanation seems to be that a symmetric expansion of the
PWN is prevented by dense material to the north of the pulsar which
shifts the whole emission to the south.  Asymmetric reverse shock
interactions of this kind have originally been proposed to explain the
offset morphology of the Vela~X PWN based on hydro-dynamical
simulations by~\citet{Blondin}. Indeed recent analyses of CO data show
dense material surrounding PSR\,J1826--1334 (at a distance of 4~kpc)
to the north and northeast~\citep{AnneCO1825}, supporting this
picture. It is interesting to note, that H.E.S.S.\ has now detected
offset morphologies from both G18.0--0.7 and Vela~X~\citep{HESSVelaX},
confirming the existence of a class of at least two offset PWN implied
by X-ray observations~\citep{XMM1825}. Whereas X-rays probe a
combination of the thermal and ultrarelativistic components, which
could have been mixed at the time when the asymmetric reverse shock
interaction took place, the H.E.S.S.\ results are important in
determining the offset morphology of the ultrarelativistic component
alone.


Based on its proximity and energetics, the pulsar PSR\,J1826--1334 has
been proposed to be associated with the unidentified EGRET source
3EG\,J1826--1302~\citep{EGRETCat}. This EGRET source exhibits a hard
power law of photon index $2.0 \pm 0.11$ with no indication of a
cut-off. The pulsar lies south of the centre of gravity of the EGRET
position and is marginally enclosed in the 95\% confidence contour
(see Fig.~\ref{fig::skymap}). It has been shown~\citep{ZhangCheng}
that an association between PSR\,J1826--1334 and 3EG\,J1826--1302 is
plausible based on the pulsar properties (such as pulsar period and
magnetic field derived in the frame of an outer gap model), and that
the observed $\gamma$-ray spectrum can be fit to this model. Although
an unpulsed excess from EGRET has been reported with a significance of
$9 \sigma$ ~\citep{NelEGRET}, a significant periodicity could not be
established. Additionally an ASCA X-ray source possibly connected to
the EGRET data above 1\,GeV~\citep{ASCAGeVRoberts} was found in this
region. Recently, \citet{NolanEGRET} reassessed the variability of the
EGRET source and found a weak variability, which led the authors to
consider the source finally as a PWN candidate in the EGRET
high-energy $\gamma$-ray energy range above 100~MeV.

Here we report on re-observations of the VHE $\gamma$-ray source
HESS\,J1825--137 and the region surrounding PSR\,J1826--1334 performed
with H.E.S.S.\ in 2005. H.E.S.S.\ consists of four imaging atmospheric
Cherenkov telescopes and detects the faint Cherenkov light from
$\gamma$-ray induced air showers in the atmosphere above an energy
threshold of 100~GeV up to several tens of TeV. Each telescope is
equipped with a mirror area of $107$~m$^2$~\citep{HESSOptics} and a
960 photo-multiplier camera for the detection of the faint Cherenkov
light. The telescopes are operated in a coincidence mode in which at
least two telescopes must have triggered in each
event~\citep{HESSTrigger}. The H.E.S.S.\ system has a point source
sensitivity above 100~GeV of $<2.0\times\,10^{-13}$
cm$^{-2}$s$^{-1}$ (1\% of the flux from the Crab nebula) for a $5
\sigma$ detection in a 25 hour observation. The system is located in
the Khomas Highland of Namibia~\citep{HESS} and began operation in
December 2003.

\section{H.E.S.S.\ observations of PSR\,J1826--1334}

First indications of a VHE $\gamma$-ray signal in the region
surrounding the pulsar PSR\,J1826--1334 during the H.E.S.S.\ Galactic
plane survey~\citep{HESSScan, HESSScanII} triggered pointed
re-observations of the region, resulting in the detection of an $8.1
\sigma$ significance signal -- named
HESS\,J1825--137~\citep{HESS1825}. This significance was obtained
using events within a circle of \emph{a priori} chosen radius $\theta
= 0.22\degr$ from the best fit position as used in a blind search for
somewhat extended sources. Using a larger integration radius of
$\theta = 0.4\degr$, appropriate to contain most of the emission
region, the significance increased to $13.4 \sigma$. HESS\,J1825--137
was reported to extend $\sim$1\degr\ asymmetrically to the south of
PSR\,J1826--1334 and shows the same asymmetric extension as the X-ray
PWN G\,18.0--0.7 on a much larger scale. In~\citet{HESS1825} an
association has been proposed between the H.E.S.S.\ source and the
X-ray emission region. The different sizes in the two energy bands
were explained by the difference in the synchrotron cooling lifetimes
of the (higher energy) X-ray emitting and the (lower energy)
IC-$\gamma$-ray emitting electrons. The energy spectrum of the source
in the 2004 data within the larger integration circle of $\theta =
0.4\degr$ was fitted by a power law of photon index
$2.40\pm0.09_{\mathrm{stat}}\pm0.2_{\mathrm{sys}}$ at a flux level
corresponding to 20\% of the flux from the Crab nebula above 1\,TeV.
The peak of HESS\,J1825--137 is located just outside the 95\%
confidence limits on the position of the unidentified EGRET source
3EG\,J1826--1302. As shown by~\citet{HESS1825}, the H.E.S.S.\ energy
spectrum can be connected to the EGRET spectrum by extrapolation. 
Therefore, despite the somewhat marginal spatial coincidence, an
association between these two objects was considered.

\begin{figure*}
  \centering
  \includegraphics[width=0.95\textwidth]{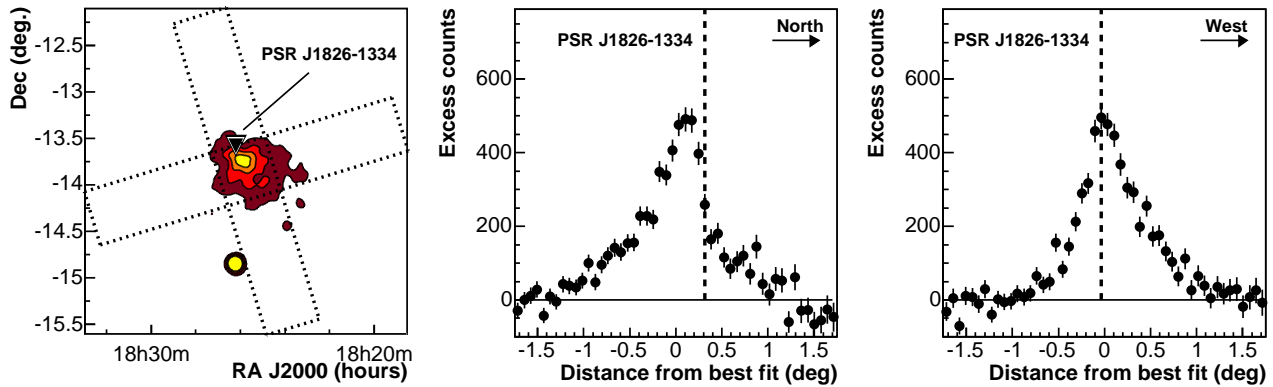}
  \caption{Slices in the uncorrelated excess map of HESS\,J1825--137
  to further illustrate the morphology. The width of the slices is
  0.6\degr\ .The direction is given by the orientation from the fit of
  an elongated Gaussian (see Fig.~\ref{fig::skymap}) and has a value
  of 17\degr\ (central panel) and perpendicular to this direction
  (right panel). The left panel shows the region in which the slices
  were taken as black boxes overlaid on the H.E.S.S.\ excess map as
  shown in Fig.~\ref{fig::skymap}. The colors denote the 20\%, 40\%,
  60\% and 80\% intensity contours of the VHE $\gamma$-ray
  emission. The slices are centred on the best fit position, the
  position of the pulsar in the slices is marked as a dashed black
  line. As a large part of the observations are taken south of
  HESS\,J1825--137, the range beyond 1\degr\ north of the pulsar
  is at the edge of the field of view of most observations, and the 
  background subtraction is less reliable in this region.}
  \label{fig::slices}
\end{figure*}

HESS\,J1825--137 was revisited in 2005 for $\sim$7 hours in pointed
observations between June and July and was additionally in the field
of view of a large part of the pointed observations on the nearby
(distance $\sim$1\degr) $\gamma$-ray emitting microquasar LS\,5039
(HESS\,J1826--146), adding another 50.9 hours between April and
September~\citep{HESS5039}. Here we report on the total available
dataset (i.e. 2004 and 2005 data) that includes now $\sim$67 hours of
observations with HESS\,J1825--137 within 2.0\degr\ of the pointing
position of the telescopes. The exposure adds up to a total dead-time
corrected lifetime of 52.1 hours after quality selection of runs
according to hardware and weather conditions, thereby increasing the
observation time by more than a factor of 6 compared to earlier
publications. The mean zenith angle of the dataset presented here is
20.1\degr, the mean offset of the peak position of HESS\,J1825--137
from the pointing direction of the system is 1.2\degr.

The standard H.E.S.S.\ event reconstruction scheme was applied to the
raw data after calibration and tail-cuts cleaning of the camera
images~\citep{HESSCalib}. The shower geometry was reconstructed based
on the intersection of the image axes, providing an angular resolution
of $\sim$0.1\degr\ for individual $\gamma$-rays. Cuts on scaled width
and length of the image (optimised on $\gamma$-ray simulations and
off-source data) are applied to select $\gamma$-ray candidates and
suppress the hadronic background~\citep{HESS2155}. The energy of the
$\gamma$-ray is estimated from the total image intensity taking into
account the shower geometry. The resulting energy resolution is
$\sim$15\%. As previously described~\citep{HESSScan, HESSRXJII}, two
sets of quality cuts are applied. For morphological studies of a
source a rather tight image size cut of 200 photo-electrons (p.e.) is
applied (along with a slightly tighter cut on the mean scaled width),
yielding a maximum signal-to-noise ratio for a hard-spectrum source.
For spectral studies the image size cut is loosened to 80~p.e.\ to
extend the energy spectra to lower energies. Different methods
are applied to derive a background estimate as described
by~\citet{HESSBack}. For morphological studies the background at each
test position in the sky is either derived from a ring surrounding
this test position (with radius 1.0\degr, an area 7 times that of the
on-source area, taking into account the changing acceptance on the
ring), or from a model of the system acceptance, derived from off-data
(data with no $\gamma$-ray source in the field of view) with similar
zenith angle. In all background methods, known $\gamma$-ray emitting
regions are excluded from the background regions to avoid $\gamma$-ray
contamination of the background estimate. All results presented here
have been obtained consistently with different background estimation
techniques.

\section{VHE $\gamma$-ray emission from HESS\,J1825--137}

To illustrate the overall morphology of HESS\,J1825--137,
Fig.~\ref{fig::skymap} shows a smoothed excess map of the field of
view surrounding the source, corrected for the changing relative
acceptance in the field of view. The background for this map has been
derived from a model of the system acceptance obtained from off-data
(similar to the background estimation in~\citet{HESSRXJII}). The map
has been smoothed with a Gaussian of width 2.5\arcmin. The inset in
the bottom left corner shows a Monte-Carlo simulated point-source as
it would appear in the same dataset taking the smoothing and the
point-spread function (PSF) for this dataset into account. The pulsar
PSR\,J1826--1334 is marked by a white triangle. To the south of
HESS\,J1825--137, another VHE $\gamma$-ray source, the point-source
microquasar LS\,5039 (HESS\,J1826--148), is
visible~\citep{HESS5039}. The color scale for this latter source is
truncated and thus its apparent size is exaggerated. Also shown in
Fig.~\ref{fig::skymap} is the 95\% positional confidence contour of
the unidentified EGRET source 3EG\,J1826--1302 (dotted white), that is
possibly associated to HESS\,J1825--137.

\begin{figure*}
  \centering
  \includegraphics[width=0.5\textwidth]{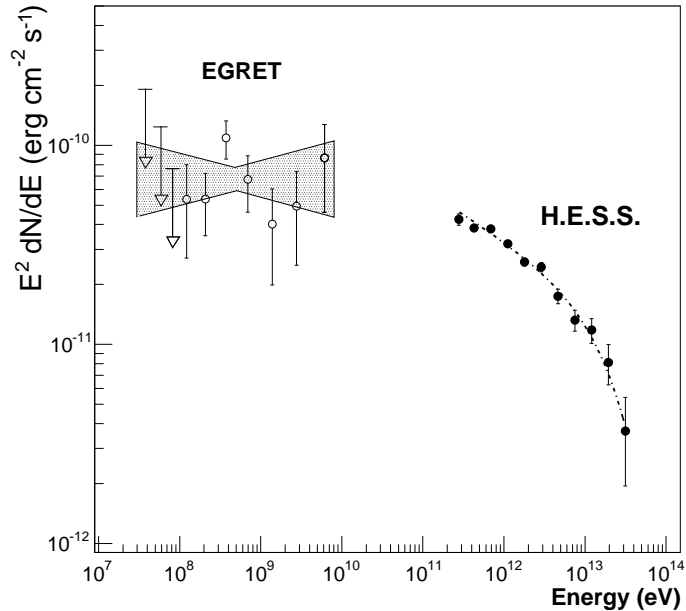}
  \caption{Energy flux $E^2dN/dE$ of HESS\,J1825--137 in $\gamma$-rays
  as measured by H.E.S.S.\ (black filled circles) up to $\sim$35~TeV
  for a large part of the emission region within an integration radius
  of 0.8\degr\ around the best fit position. Also shown as open
  circles are the energy flux points of the possibly related
  unidentified EGRET source 3EG\,J1826--1302. The background estimate
  for the spectrum has been derived from off-runs (i.e. runs without
  $\gamma$-rays in the field of view). The spectrum can be fit by a
  power law $dN/dE = I_0\ E ^ {-\Gamma}$ but the poor $\chi^2$ of the
  fit (40.4 for 15 degrees of freedom) suggests a different spectral
  shape. Here a power law fit with an exponential cutoff is shown, but
  it should be noted that this shape is not favoured over other
  similarly complex spectral shapes as discussed in the text in more
  detail.
  }
  \label{fig::sed} 
\end{figure*}

HESS\,J1825--137 shows a clearly extended morphology with respect to
the PSF, extending to the south-west of the pulsar.  The position and
extension of HESS\,J1825--137 have been determined by fitting the
uncorrelated (i.e. unsmoothed) excess map to a model of a 2-D Gaussian
$\gamma$-ray brightness profile of the form $\rho \propto
\exp(-\theta^{2}/2\sigma_{\mathrm{source}}^2)$, convolved with the PSF
for this dataset (68\% containment radius: 0.075\degr). The best fit
position -- equivalent to the center of gravity of the source -- is at
18h25m41s $\pm 3_{\mathrm{stat}}$s , --13\degr50\arcmin21\arcsec $\pm
35_{\mathrm{stat}}$\arcsec (here and in the following the epoch J2000
is used), the best fit rms extension is $\sigma_{\mathrm{source}} =
0.24\degr \pm 0.02_{\mathrm{stat}}\degr$ .  However, the $\chi^2$ per
degree of freedom is not satisfactory ($1295/1085$), indicative of the
more complex morphology of the source.  Reflecting the non-Gaussian
and skewed source profile, the position of the peak in the
$\gamma$-ray emission (at 18h25m57s, --13\degr43\arcmin36.8\arcsec as
determined by fitting a 2-D Gaussian in a restricted region around the
peak) is slightly shifted at a distance of $\sim$8\arcmin\ to the best
fit position. The pulsar PSR\,J1826--1334 is located at a distance of
$\sim$10\arcmin\ from the peak $\gamma$-ray emission and
$\sim$17\arcmin\ from the best fit position.  To test for a different
source morphology, an elongated Gaussian with independent
$\sigma_{\mathrm{source}}$ along the minor and
$\sigma^{\prime}_{\mathrm{source}}$ along the major axis and a free
position angle $\omega$ (measured counter-clockwise from the North)
has also been fitted. This elongated fit gives a best fit position of
18h25m41s $\pm 4_{\mathrm{stat}}\mathrm{s}$,
--13\degr50\arcmin20\arcsec $\pm 40_{\mathrm{stat}}\arcsec$,
consistent within errors to the symmetrical fit. The fit yields only a
slight indication for an elongation with $\sigma_{\mathrm{source}} =
0.23 \pm 0.02_{\mathrm{stat}}$ and $\sigma^{\prime}_{\mathrm{source}}
= 0.26 \pm 0.02_{\mathrm{stat}}$ at a position angle of $\omega =
17\degr \pm 12\degr_{\mathrm{stat}}$. The $\chi^2$ per degree of
freedom ($1288/1083$) is still relatively poor.  The best fit position
deviates slightly from the best fit position reported in earlier
papers \citep{HESSScan, HESSScanII}.  The difference can mainly be
attributed to the different fit range. The best fit parameters of the
elliptical fit are shown as a black square and ellipse in
Fig.~\ref{fig::skymap}. Note that the fitted position angle is
consistent within errors with the orientation of the line connecting
the pulsar position and the best fit position, which amounts to
23.1\degr.

Fig.~\ref{fig::slices} shows slices in the direction of the position
angle (17\degr) of the elliptical fit (centre) and in the direction
perpendicular to it (right). The width of the slices is chosen to be
0.6\degr, the slices are illustrated in the left panel as black dashed
boxes. The position of the pulsar in the slices is marked as a dashed
black line. It can be seen, that the peak of the H.E.S.S.\ emission is
close to the pulsar position but slightly shifted as is also apparent
from the two-dimensional excess plot. Also visible in the central
panel is the rather sharp drop from the peak position towards the
north-eastern direction and the longer tail to the south-western
direction. Some indication for an additional excess to the north of
HESS\,J1825--137 is seen in Fig.~\ref{fig::skymap} and in the
central panel of Fig.~\ref{fig::slices} at a distance of
$\sim$0.7\degr\ from the pulsar position. Further investigation of
this feature will have to await future data, in particular given that
most current data were taken on positions south of the pulsar, with
regions in the north near the edge of the field of view.

\begin{table*}
    \caption{Fit results for different spectral models for the whole
      emission region within an integration radius of 0.8\degr\ around
      the best fit position and the background derived from
      off-data. The differential flux normalisation $I_0$ is given in
      units of $10^{-12} \, \mathrm{cm}^{-2} \, \mathrm{s}^{-1} \,
      \mathrm{TeV}^{-1}$. $E$, $E_{\mathrm{B}}$, and $E_{\mathrm{c}}$
      are given in units of TeV. The last column gives the integrated
      flux above the spectral analysis threshold of 270~GeV in units
      of $10^{-11}\, \mathrm{cm}^{-2} \, \mathrm{s}^{-1}$. The
      power-law fit provides a rather poor description of the
      data. Thus fits of a power law with an exponential cutoff (row
      2), a power law with an energy dependent photon index (row 3),
      and a broken power law (row 4; in the formula, the parameter $S
      = 0.1$ describes the sharpness of the transition from $\Gamma_1$
      to $\Gamma_2$ and is fixed in the fit) are also given. Note that
      some of the fit parameters are highly correlated.}
    \label{tab::fits}
    \centering
    \begin{tabular}{|l| l l l l | l | l |}
      \hline
      Fit Formula for $\frac{\mathrm{d} N}{\mathrm{d} E}$& 
      \multicolumn{3}{c}{Fit Parameters}  & & $\chi^2$~(ndf) &
      Flux$_{> 270~\mathrm{GeV}}$\\
      \hline
      \hline
      $\displaystyle{I_0\ E ^ {-\Gamma}}$ & $I_0 = 19.8 
      \pm 0.4$ & $\Gamma = 2.38 \pm 0.02$ & & &
      40.4 (15) & $87.4 \pm 2.0$ \\
      $\displaystyle{I_0\ E ^ {-\Gamma}\ \exp (-E / E_{\mathrm{c}})}$ & $I_0 = 21.0
      \pm 0.5$ & $\Gamma = 2.26 \pm 0.03$ & $E_{\mathrm{c}} = 24.8 \pm
      7.2 $ & & 16.9 (14) & $86.7 \pm 2.5$\\
      $\displaystyle{I_0\ E ^ {-\Gamma + \, \beta\ \log E}}$ & $I_0 = 21.0
      \pm 0.4$ & $\Gamma = 2.29 \pm 0.02$ & $\beta = -0.17 \pm
      0.04 $ & & 14.5 (14) & $82.8 \pm 2.2$ \\
      $\displaystyle{I_0\ \left( E / E_{\mathrm{B}} \right) ^ {-\Gamma_1}\ \left( 1 + \left(E /
      E_{\mathrm{B}}\right) ^ {1 / S } \right) ^ {\, S \, (\Gamma_1
      - \Gamma_2)}}$ & $I_0 = 2.2 \pm 1.0$ & $\Gamma_1
      = 2.26 \pm 0.03$ & $\Gamma_2 = 2.63 \pm 0.07$ & $ E_{\mathrm{B}} =
      2.7 \pm 0.5 $ & 15.1 (13) & $84.6 \pm 38.5 $\\[0.15cm]
      \hline
    \end{tabular}
\end{table*}

For the spectral analysis the image size cut is loosened to 80~p.e.\
to achieve a maximum coverage in energy. The resulting spectral
analysis threshold for the dataset described here is 270~GeV. Events
with reconstructed direction within an angle $\theta = 0.8\degr$ of
the source location are considered on-source.  No correction for the
$\gamma$-ray emission extending beyond this angular cut has been
applied. Thus the flux level determined corresponds to the flux level
of the source within the integration region and might be an
underestimation of the flux from the whole source. In the
determination of the energy spectrum, the energy of each event is
corrected for the time-varying detector optical efficiency, relative
to that used in Monte Carlo simulations to estimate the effective area
of the instrument. The optical efficiency is estimated from single
muon events detected during each observation run
\citep{leroy03,bolz04}. The mean energy correction is $\sim$$25\%$.
For the spectral analysis the background is taken from positions in
the same field of view with the same offset from the pointing
direction as the source region. This approach is taken to avoid
systematic effects from the energy-dependent system acceptance
function (which is to a good approximation radially symmetric). In
another approach off-data have been used in the background estimation
to confirm the results from the same field of view, using either the
same shaped region as the on-region in the off-data or using again
off-regions distributed with the same offset from the pointing
direction of the system as the on-region. The total significance of
the emission region with the loose cuts is $33.8 \sigma$ with an
excess of $19510 \pm 577$ $\gamma$-ray events. Fig.~\ref{fig::sed}
shows the spectral energy distribution in terms of energy flux
$E^2dN/dE$ of the H.E.S.S.\ emission region (full black circles). Also
shown are the energy flux points and the spectral fits of the possibly
related unidentified EGRET source 3EG\,J1826--1302 (open
circles). Given the poor angular resolution of EGRET, these data are
taken on a scale similar to that of the full H.E.S.S.\ emission region
and can thus be compared to the total H.E.S.S.\ flux. From this figure
one can see that the unidentified EGRET source 3EG\,J1826--1302 could
be associated with the H.E.S.S.\ emission region from a spectral
continuity point of view.

\begin{figure*}
  \centering
  \includegraphics[width=0.6\textwidth]{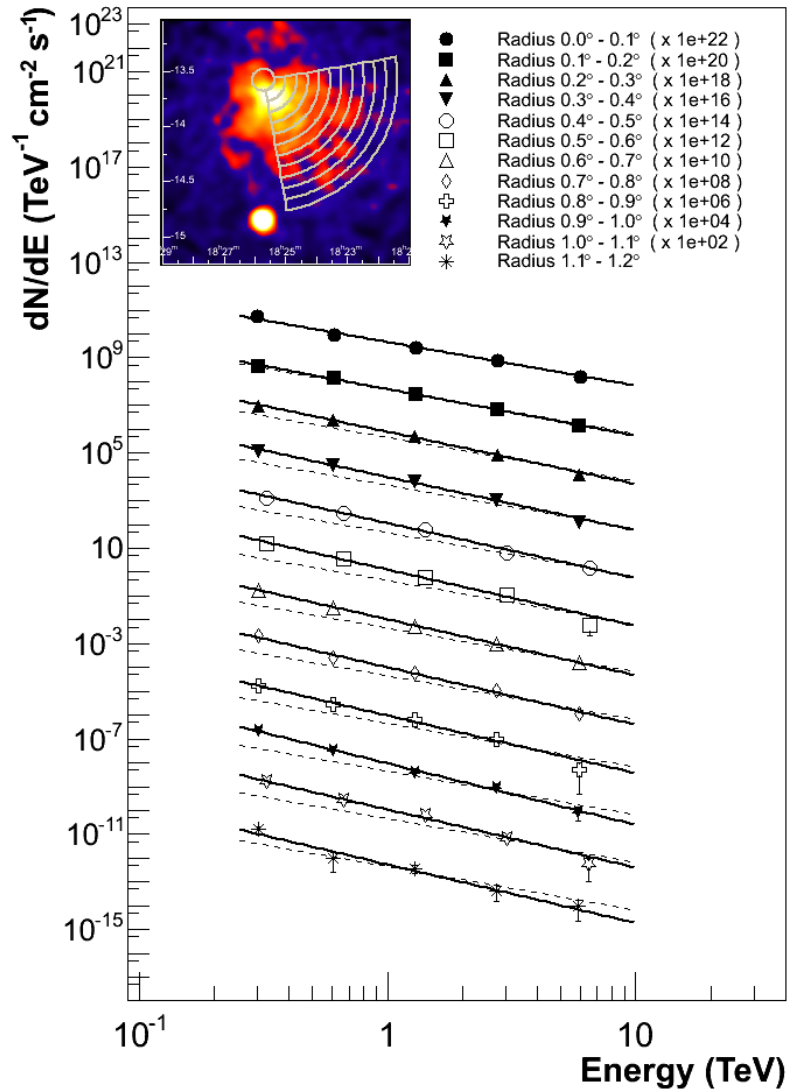}
  \caption{Energy spectra in radial bins.  {\bf Inset:} H.E.S.S.\
  excess map as shown in Fig.~\ref{fig::skymap}. The wedges show the
  radial regions with radii in steps of 0.1\degr\ in which the energy
  spectra were determined. The innermost region is centred on the
  pulsar PSR\,J1826--1334. {\bf Main Figure:} Differential energy
  spectra for the regions illustrated in the inset, scaled by
  powers of 10 for the purpose of viewing. The spectrum for the
  analysis at the pulsar position is shown as a reference along with
  the other spectra as dashed line. For all regions the energy
  spectrum has been determined as described in the text and has been
  fitted by a power-law in a restricted energy range between 0.25 and
  10 TeV.}
  \label{fig::spectraradialAll}
\end{figure*}

A fit of the differential energy spectrum from 270~GeV up to
$\sim$35~TeV by a power law $dN/dE = I_0\ E ^ {-\Gamma}$ yields a
normalisation of $I_0 = 19.8 \pm 0.4_{\mathrm{stat}}\pm
4.0_{\mathrm{sys}} \times 10^{-12}$ TeV$^{-1}$ cm$^{-2}$ s$^{-1}$ and
a photon index $\Gamma =
2.38\pm0.02_{\mathrm{stat}}\pm0.15_{\mathrm{sys}}$ (see
Table~\ref{tab::fits}). The flux of HESS\,J1825--137 above 1\,TeV
corresponds to $\sim$68\% of the flux from the Crab nebula. Note that
this flux is significantly higher than the previously reported
flux~\citep{HESS1825} due to a significantly increased integration
radius (0.8\degr\ instead of 0.4\degr) in the attempt to cover the
whole source region. Integrating only within the smaller region of
0.4\degr\ the flux level is consistent with the previously published
result. The power-law fit represents a rather bad description of the
data (as can be seen $\chi^2$ of the fit) and suggests therefore a
different spectral shape. Various models have been fit to the data to
investigate the shape of the spectrum. Table~\ref{tab::fits}
summarises these fits. Three alternative shapes have been used: a
power law with an exponential cutoff $E_{\mathrm{c}}$ (row 2), a power
law with an energy dependent exponent (row 3), and and a broken power
law (row 4).
In all cases, $I_0$ is the differential flux normalisation, and the
photon indices are specified as $\Gamma$. It is evident that the
alternative descriptions of the spectrum describe the data
significantly better than the pure power law as can be seen from the
decreasing $\chi^{2}/\mathrm{ndf}$ (see Table~\ref{tab::fits}).

\begin{figure*}
  \centering
  \includegraphics[width=0.8\textwidth]{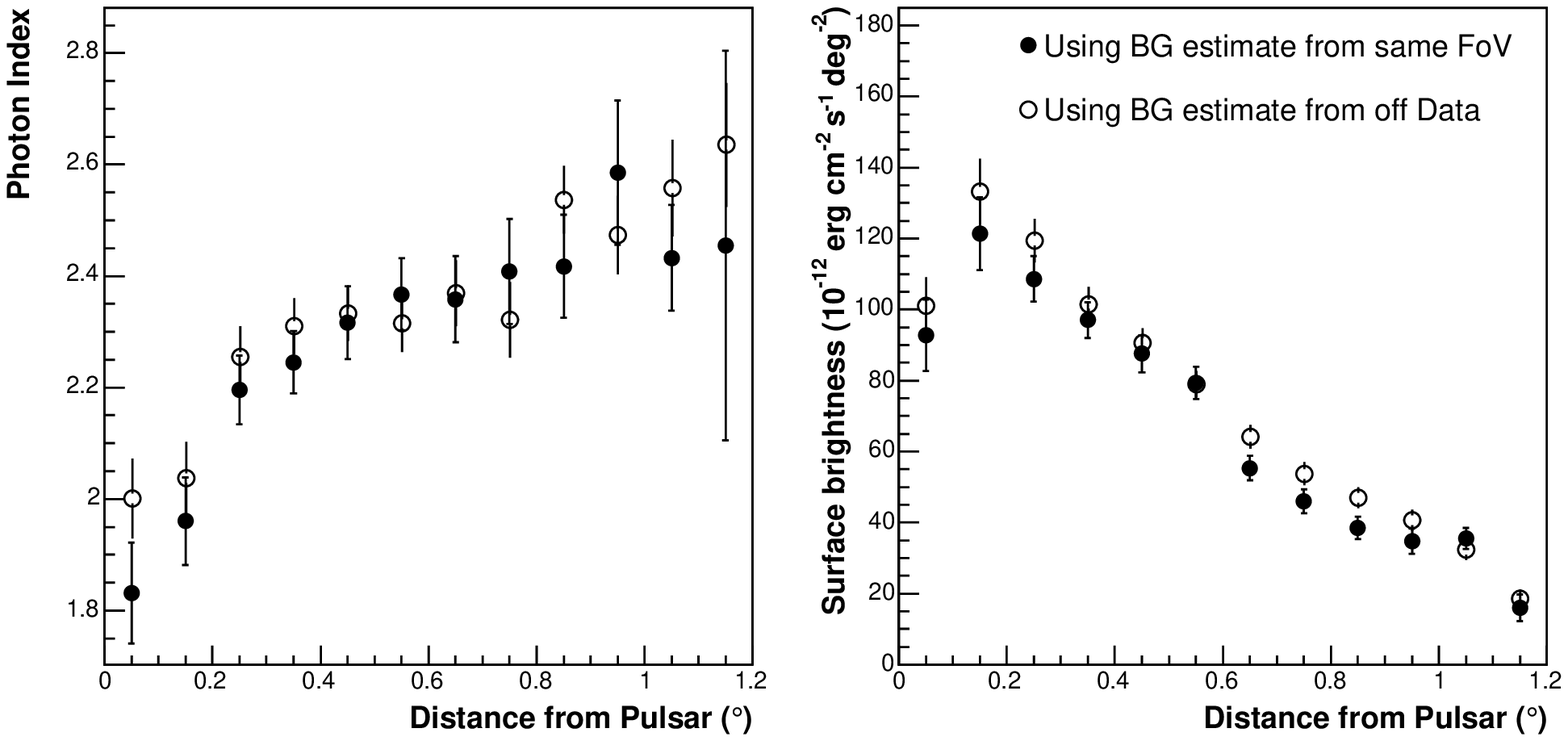}
  \caption{Energy spectra in radial bins.  {\bf Left:} Power-law
  photon index as a function of the radius of the region (with respect
  to the pulsar position) for the regions given in
  Fig.~\ref{fig::spectraradialAll}. The closed points are obtained
  by deriving the background estimate from regions with the same
  offset as the on-region within the same field of view. The open
  points are derived using off-data (data without $\gamma$-ray sources
  in the field of view) for the background estimate. A constant fit to
  the filled points yields a $\chi^2$ of 58 for 11 degrees of freedom,
  clearly showing the dependence of the photon index on the distance
  from the pulsar position. 
{\bf Right:} Surface brightness between 0.25 and 10 TeV per
  integration region area in units of $10^{-12} \mathrm{erg} \,
  \mathrm{cm}^{-2} \, \mathrm{s}^{-1} \, \mathrm{deg}^{-2}$ as a
  function of the distance to the pulsar position. Error bars denote
  $\pm 1 \sigma$ statistical errors. }
  \label{fig::spectraradial}
\end{figure*}
\begin{table*} 
  \caption{Spectral parameters for the radial bins surrounding
  PSR\,J1826--1334. \emph{PSF} denotes a H.E.S.S.\ point-source
  analysis at the pulsar position. The background estimate for the
  numbers in the table have been derived from reflected positions
  within the same field of view. The energy flux and surface
  brightness are given for the energy range between 0.25 and 10~TeV.}
  \label{tab::energetics}
  \centering
  \begin{tabular}{| r | c  c  c  c  c |}\hline 
    Radius & Photon Index & $\chi^{2}$/ndf & Area & Energy Flux & Surface
    Brightness \\
    (deg) & & & (deg$^{2})$ & (erg cm$^{-2}$ s$^{-1}$) & (erg cm$^{-2}$
    s$^{-1}$ deg$^{-2}$)\\
    \hline
    \hline
    PSF & 1.83 $\pm$ 0.09 & 2.8 / 3 & 3.1e-02 & 2.9e-12 $\pm$ 3.2e-13 & 9.3e-11 $\pm$ 1.0e-11\\
    0.15 &1.96 $\pm$ 0.08 & 0.8 / 3 & 2.4e-02 & 2.9e-12 $\pm$ 2.4e-13 & 1.2e-10 $\pm$ 1.0e-11\\
    0.25 &2.20 $\pm$ 0.06 & 3.1 / 3 & 3.9e-02 & 4.3e-12 $\pm$ 2.5e-13 & 1.1e-10 $\pm$ 6.4e-12\\
    0.35 &2.25 $\pm$ 0.06 & 6.9 / 3 & 5.5e-02 & 5.3e-12 $\pm$ 2.8e-13 & 9.7e-11 $\pm$ 5.1e-12\\
    0.45 &2.32 $\pm$ 0.07 & 7.1 / 3 & 7.1e-02 & 6.2e-12 $\pm$ 3.7e-13 & 8.8e-11 $\pm$ 5.2e-12\\
    0.55 &2.37 $\pm$ 0.06 & 8.5 / 3 & 8.6e-02 & 6.9e-12 $\pm$ 3.9e-13 & 7.9e-11 $\pm$ 4.5e-12\\
    0.65 &2.36 $\pm$ 0.08 & 0.4 / 3 & 1.0e-01 & 5.7e-12 $\pm$ 3.6e-13 & 5.5e-11 $\pm$ 3.5e-12\\
    0.75 &2.41 $\pm$ 0.09 & 8.3 / 3 & 1.2e-01 & 5.4e-12 $\pm$ 4.0e-13 & 4.6e-11 $\pm$ 3.4e-12\\
    0.85 &2.42 $\pm$ 0.09 & 6.0 / 3 & 1.3e-01 & 5.1e-12 $\pm$ 4.2e-13 & 3.8e-11 $\pm$ 3.1e-12\\
    0.95 &2.59 $\pm$ 0.13 & 2.4 / 3 & 1.5e-01 & 5.2e-12 $\pm$ 5.4e-13 & 3.5e-11 $\pm$ 3.6e-12\\
    1.05 &2.43 $\pm$ 0.09 & 6.4 / 3 & 1.6e-01 & 5.9e-12 $\pm$ 4.9e-13 & 3.6e-11 $\pm$ 3.0e-12\\
    1.15 &2.45 $\pm$ 0.35 & 3.4 / 3 & 1.8e-01 & 2.9e-12 $\pm$ 6.8e-13 & 1.6e-11 $\pm$ 3.8e-12\\
\hline
\end{tabular}
\end{table*}

Given the large dataset with more than 19,000 $\gamma$-ray excess
events and given the extension of HESS\,J1825--137, a spatially
resolved spectral analysis has been performed to search for a change
in photon index across the source, similar to the detailed analysis of
the $\gamma$-ray SNR RX\,J1713.7--3946 as performed
in~\citet{HESSRXJII}. Fig.~\ref{fig::spectraradialAll} shows energy
spectra determined in radial bins around the pulsar position, covering
the extended tail of the VHE $\gamma$-ray source.  The inset of
Fig.~\ref{fig::spectraradialAll} shows again the H.E.S.S.\ excess map
as shown in Fig.~\ref{fig::skymap} along with wedges that
illustrate the regions in which the energy spectra were determined,
with radii increasing in steps of 0.1\degr; the innermost region is
centred on the pulsar PSR\,J1826--1334.  The opening angle of the
wedges was constrained by LS\,5039 in the southern part and by the
apparent end of the emission region in the northern part. For all
regions the energy spectrum has been determined by defining the wedge
as the on-region. The background estimate has been derived from
circles distributed on a ring around the pointing direction. The
radius of this ring was chosen to be equal to the distance of the
centre of gravity of the wedge to the pointing direction. This
approach ensures a similar offset distribution in the on- and
off-dataset and has been used to determine the background estimate
from the same field of view as well as from off-data taken on regions
without $\gamma$-ray sources.
Consistent results were achieved in both methods.

Along with each spectrum in Fig.~\ref{fig::spectraradialAll}, the
power law fit to the innermost region centred on the pulsar position
is shown as a dashed line for comparison. A softening of the energy
spectra is apparent with increasing distance from the pulsar.  This
softening is equivalent to a decrease of the source size with
increasing energy and provides the first evidence for an energy
dependent morphology detected in VHE $\gamma$-rays. Differences in
the energy bin sizes arise from the fact that for non-significant
photon points the bin size was increased. It has been verified that
this approach does not change the result of the fit. Due to the
different distribution of offsets from the pointing direction of the
system in the different regions, the photon analysis threshold changes
slightly, thus some of the different spectra do not start at exactly
the same energy.

Fig.~\ref{fig::spectraradial} summarises the findings of
Fig.~\ref{fig::spectraradialAll} by plotting the fit parameters of the
power law fit versus the distance of the region to the pulsar
position. Shown are the results using two different background
estimation techniques in the spectral analysis. The left panel shows
the photon index as a function of the distance from the pulsar. A
clear increase of the photon index for larger distances from the
pulsar position is apparent; the photon index seems to level off
within errors to a value of $\sim$2.4 $\pm 0.1$ at a distance of
$\sim$0.6\degr. The right panel shows the surface brightness (i.e. the
integrated energy flux $E dN/dE$ per unit area between 0.25~TeV and
10~TeV) as a function of the distance to the pulsar position. Again
here it can be seen, that the maximum of the emission is slightly
shifted away from the pulsar position as was already apparent in
Fig.~\ref{fig::slices}. In both panels, the error bars denote $\pm 1
\sigma$ statistical errors. Systematic errors of 20\% on the flux and
0.15 on the photon index are to be assigned to each data point in
addition. However, since all spectra come from the same set of
observations, these systematic errors should be strongly correlated,
and will cancel to a large extent when different wedges are
compared. Table~\ref{tab::energetics} summarises the different
spectral parameters determined in the wedges using the reflected
background from the same field of view.

Whereas the H.E.S.S.\ observation of an energy dependent morphology
represents the first detection of such an effect in $\gamma$-ray
astronomy, the dependence of observed photon index on radius (commonly
known as "$\Gamma-r$" relation) is well known from X-ray observations
of PWN other than the Crab. For \object{G21.5--0.9}~\citet{SlaneG21.5}
found a value of $\Gamma \sim1.5$ near the PWN termination shock,
after which it converges to a value of $\sim$2.2 in the outer
nebula. For the PWN \object{3C58}, the photon index increases from 1.9
in the torus to $\sim$2.5 at the edge of 3C58~\citep{Slane3C58}. For
this object~\citet{Bocchino3C58}, also found that the energy flux per
radial interval for 3C58 remains approximately constant, consistent
with our findings for HESS\,J1825--137.  A similar constant energy
flux with increasing radius was also found in X-ray observations of
another VHE $\gamma$-ray source \object{G0.9+0.1}~\citep{PorquetG09}.
For this composite remnant, the photon index also varies with radius
from 1.5 (beyond the compact core) to $\sim$2.5 near the edge of the
PWN. In the case of the more evolved Vela PWN~\citet{ManganoVela}
found a radial variation of 1.55 to 2.0. For most of these remnants a
total change in the photon index of $\sim$0.5 is seen, as expected for
cooling losses. Attempts to model the $\Gamma-r$ relationship were not
successful in the past -- \citet{Slane3C58} showed that
the~\citet{Kennel1984} model for convective flow (which includes the
conservation of magnetic flux) fails to reproduce this well-known
$\Gamma-r$ relationship for PWN which are evolved beyond the Crab
phase. It should additionally be noted, that the H.E.S.S.\ observation
of an energy dependent morphology is the first unambiguous detection
of a spectral steepening away from the pulsar, for fixed electron
energies; in X-rays the situation is complicated by a possible
variation of the magnetic field with increasing distance from the
pulsar; if the X-ray spectrum is probed near or above the peak of the
SED, a variation of the field will influence the slope. Depending on
the age and magnetic field, one might expects to see similar effects
in other VHE $\gamma$-ray PWN, but so far only HESS\,J1825--137 has
sufficient statistics to clearly reveal the energy dependent
morphology.

\begin{figure*}
  \centering
  \includegraphics[width=0.9\textwidth]{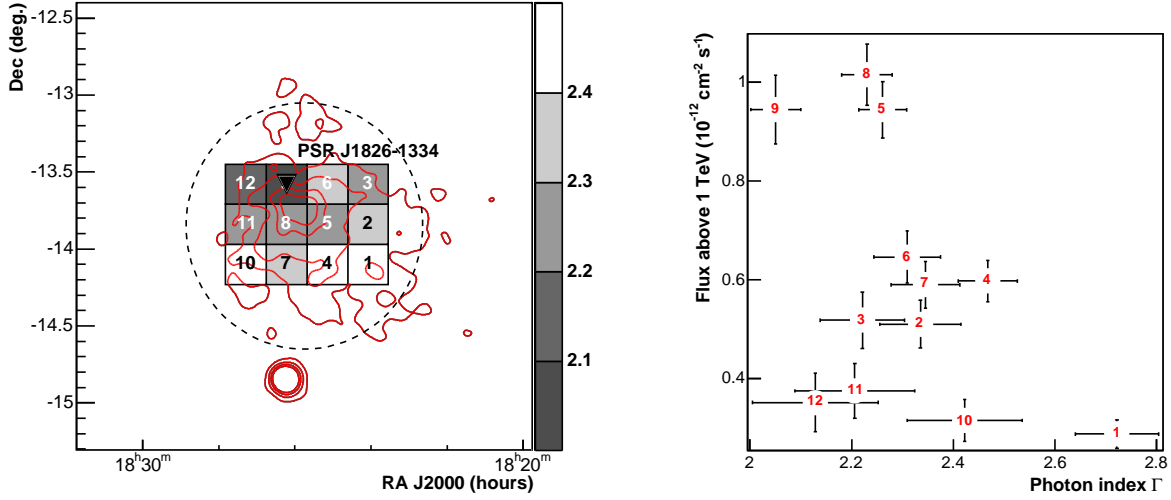}
  \caption{Spatially resolved spectral analysis of
    HESS\,J1825--137. \textbf{Left:} Shown in solid red are VHE
    $\gamma$-ray excess contours from Fig.~\ref{fig::skymap}, linearly
    spaced at the 20\%, 40\%, 60\% and 80\% maximum intensity
    levels. Superimposed are the 12 boxes ($0.26\degr \times
    0.26\degr$) for which spectra were obtained. The black dashed line
    is the $0.8\degr$ radius circle that was used to integrate events
    to produce a spectrum of the whole region. The photon index
    obtained from a power-law fit in each region is colour coded in
    bins of 0.1. \textbf{Right:} Integral flux above 1\,TeV per square
    degree versus photon index, for the 12 regions shown in the left
    panel. The error bars are $\pm 1 \sigma$ statistical errors. The
    correlation coefficient for the two quantities is $-0.46 \pm
    0.14$.}
  \label{fig::spectraboxes}
\end{figure*}

To further investigate the spectral properties of HESS\,J1825--137,
the emission region has been segmented into boxes. The result of the
spectral analysis in these boxes is shown in
Fig.~\ref{fig::spectraboxes}. The left panel shows in red VHE
$\gamma$-ray excess contours as given in
Fig.~\ref{fig::skymap}. Overlaid are 12 boxes for which spectra were
obtained independently.
The photon index resulting from a power law fit in each region is
grey-scale coded in bins of 0.1. Also here a softening of the spectral
indices away from the pulsar position is apparent, although the error
bars are larger than in Fig.~\ref{fig::spectraradial} due to the
smaller integration regions. The size of the boxes is equivalent to
the ones used in the analysis of the shell-type SNR
RX\,J1713.7--3946~\citep{HESSRXJII}, where no spectral variation has
been detected. The right hand figure shows the correlation of photon
index $\Gamma$ to integral flux per square degree above 1\,TeV. A mild
correlation between the flux per deg$^2$ and the spectral index exists
and the correlation coefficient between these two quantities is $-0.46
\pm 0.14$.

\section{Interpretation}
Obviously, the key aspect of the new H.E.S.S.\ data is the softening
of the energy spectrum at larger distances from the pulsar position or
equivalently, the decrease in source size with increasing energy of
the $\gamma$-rays. The following discussion is based on the plausible
assumption that particles have been accelerated in the proximity of
the pulsar during the last 20 kyears -- the characteristic age of the
pulsar -- and that they were then transported by diffusion and/or
convection to form the extended nebula of about 10--15~pc radius
\citep[for a recent review of PWN see][]{GaenslerReview}. The
asymmetric shape of the nebula most likely reflects the ambient
conditions, and could be caused by a reverse supernova shock created
in the dense medium north of the pulsar~\citep{Swaluw, HESS1825},
geometrically by the motion of the pulsar leaving a PWN `trail', or
even dynamically by a bow shock, resulting in a `cometary' tail;
however, the high speeds in excess of 1000~km/s required in case of
HESS J1825--137 favour the first scenario.  Spectral variation with
distance from the pulsar could result from (i) energy loss of
particles during propagation, with radiative cooling of electrons
propagating outward from the pulsar termination shock as the obvious
loss mechanism, as well as from (ii) energy dependent diffusion or
convection speeds, and from (iii) variation of the shape of the
injection spectrum with age of the pulsar which, after propagation,
translates into a spatial variation of spectra.

Loss mechanisms in (i) include, e.g, adiabatic expansion, ionisation
loss, bremsstrahlung, synchrotron losses and inverse Compton (IC)
losses; only the last two result in a lifetime $\tau =
E/(\mbox{d}E/\mbox{d}t)$ which decreases with energy and hence causes
power-law spectra to steepen, due to the quadratic dependence of
$\mbox{d}E/\mbox{d}t$ on the particle energy~\citep{Blumenthal,
Kardashev}. A source size which decreases with energy is therefore a
strong indication that the accelerated particles are electrons.  The
lifetime due to synchrotron and IC losses is:
\begin{equation}
\tau_{\rm rad} \approx 3.1  \times 10^5 
\left(\frac{w_r}{\rm eV\,cm^{-3}}\right)^{-1}
\left(\frac{E_e}{\rm TeV}\right)^{-1}\,\rm yr. \ 
\label{liftime}
\end{equation}  
Here, $w_r=\eta w_{\rm ph} + 0.025\, \rm eV cm^{-3} (\mathrm{B}/
\mu\rm G)^2 $ is the total energy density in the form of radiation and
magnetic fields, $\eta$ is a normalisation factor to account for the
reduction of IC losses due to the transition into the Klein-Nishina
regime and $B$ the magnetic field.  Given the density of the 2.7~K
CMBR $w_{\rm 2.7 K}=0.26 \ \rm eV cm^{-3}$, and the fact that IC
losses of multi-TeV electrons on the diffuse optical/IR photons are
strongly suppressed due to the Klein-Nishina effect, and even for the
CMBR are reduced by a factor $\eta \approx 2/3$ at H.E.S.S.\ energies,
synchrotron losses dominate for $B > 3~\mu$G.  In case of continuous
injection and radiative lifetimes short compared to the age of the
source, $\tau_{\rm rad}(E) \ll T$, the spectral index $\alpha$ of the
electrons steepens by one unit, corresponding to a change of the
photon index by half a unit, which approximately matches the observed
variation between the inner and outer regions of the nebula
(Fig.~\ref{fig::spectraradial}). In the Thomson regime, the energy of
the parent electrons is $E_e \approx 20 (E_\gamma / \rm TeV \ )^{1/2}
\rm \, TeV $, corresponding to the range from about 10 to 100~TeV for
$\gamma$-ray energies between 0.2 and 20~TeV.  Cooling time scales
below 20~kyears require $B > 6~\mu$G for 10~TeV electrons; at 100~TeV
the lifetime is below 20 kyears already in typical 3~$\mu$G
interstellar fields, so some steepening of spectra at the highest
energies is expected even in relatively modest fields.

It is then instructive to consider the energy budget of the PWN in an
electronic scenario.  The assumed large distance of $\approx 4$~kpc
and the relatively high $\gamma$-ray flux, $F \simeq 1.5 \times 10^{-10}
\ \rm erg/cm^2 s$ above 200~GeV, imply a quite luminous VHE
$\gamma$-ray source, $L_\gamma \sim 3 \times 10^{35} \ \rm erg/s$. This
luminosity is comparable to that of the Crab nebula, while the
spin-down luminosity of the pulsar is smaller by two orders of
magnitude. Thus, the efficiency of the $\gamma$-ray production in
HESS\,J1825--137 is much higher, $\epsilon_\gamma = L_\gamma (> 200
\rm GeV) /L_{\mathrm{rot}} \approx 0.1$.  A relatively large efficiency is
not unexpected~\citep{AhaAtoKif97} since the much lower magnetic field
in a nebula powered by a less energetic pulsar results in a more
favourable sharing between IC and synchrotron energy losses.  In a
steady state, and neglecting non-radiative energy losses, the
efficiency for $\gamma$-ray production is
\begin{equation}
\epsilon_\gamma \approx \epsilon_e {\tau_{\rm rad} \over \tau_{\rm
    IC}} \approx   
\epsilon_e {\eta w_{\rm ph} \over \eta w_{\rm ph} + 0.025 \rm eV cm^{-3}
    (\mathrm{B}/ \mu\rm G)^2} 
\end{equation}
where $\tau_{\rm IC}$ is the lifetime due to IC losses and
$\epsilon_e$ is the fraction of pulsar spin-down power going into 10
to 100~TeV electrons, corresponding to the observed $\gamma$-ray
energies.  Unless the electron spectral index is well below 2 at
energies below 10~TeV, $\epsilon_e$ will be below 10-15\%, taking into
account the sharing of spin-down energy between particle and field
energies. An efficiency $\epsilon_\gamma$ of 0.1 cannot be obtained,
even for rather small magnetic fields in the range of a few $\mu$G.
Detailed numerical simulations with (optionally time-dependent)
electron injection and cooling confirm that an energy input about one
order of magnitude higher than the current spin-down luminosity is
required to sustain the observed gamma-ray flux and to quantitatively
reproduce the measured spectrum, assuming that the distance of
$\sim4$~kpc is correct. A likely solution is that the spin-down power
of the pulsar was significantly higher in the past; for modest $B$
fields of a few $\mu$G electron lifetimes in particular at lower
energies are of the order of the pulsar age and the time variation of
spin-down luminosity needs to be taken into account. For example, with
$L_{\mathrm{rot}} \propto t^{-2}$ for a braking index of $n=3$,
`relic' electrons released in the early history of the pulsar and
surviving until today can provide sufficient energy.  To allow
accumulation of electrons over the history of the pulsar, magnetic
fields should not exceed 10\,$\mu$G.

A discussion of the energy-dependent morphology requires assumptions
concerning the transport mechanism. At least in the inner regions of
the nebula, convection is likely to dominate over diffusion.  Indeed,
the variation of surface brightness across the source -- roughly
proportional to $1/\theta$, where $\theta$ is the angular distance
from the pulsar (see Table~\ref{tab::energetics}) -- is difficult to
account for in purely diffusive propagation.  A surface brightness
$\propto \theta^{-n}$ is obtained -- for spherical symmetry -- from a
volume density $\propto r^{-n-1}$. Neglecting cooling effects, a
$1/\theta$ dependence is hence obtained for a constant radial
convection velocity, resulting in a $1/r^2$ density distribution.  For
constant convection speed, energy conservation requires a rapid
decrease of $B$-fields with distance from the pulsar, with very low
fields at the edge of the PWN unless one is dealing with a very strong
and young source such as the Crab nebula~\citep{Kennel1984}. A
convection speed $v(r) \propto 1/r$ would allow a constant
$B$-field. Such convection results in constant surface density;
however, the electron density at a fixed electron energy -- and
therefore the $\gamma$-ray intensity -- will again decrease with
distance once cooling is included. A speed $v(r) \propto 1/r$ results
in a propagation time $t \sim r^2$ and, at energies where the electron
lifetime $\tau_{\rm rad} \propto 1/E_e$ is shorter than the lifetime
$T$ of the accelerator, in a source size $R \propto E_e^{-1/2}$.

A similar result is obtained for the diffusion case (ii), which is
expected to be relevant near the outer edge of the nebula. The
diffusive source size is governed by the diffusion coefficient $D(E)$,
which is frequently parametrised in a power-law form $D(E)=D_0
(E/E_0)^\delta$, with $\delta$ between 0 for energy-independent
diffusion and 1 for Bohm diffusion. The resulting size can be
estimated to $R \simeq [2 D(E) t]^{1/2}$ with the propagation time $t$
again given by the age $T$ of the accelerator or the lifetime
$\tau_{\rm rad}$ of radiating particles, whatever is smaller. For
lifetimes $\tau_{\rm rad} \ll T$ short compared to the age of the
accelerator, one obtains $R \propto E_e^{(\delta-1)/2}$.  In case of
Bohm-type diffusion with $\delta=1$, the radiative losses and the
diffusion effects compensate each other and the size becomes
effectively energy independent. For energy independent diffusion,
i.e. $\delta=0$, the size decreases with energy again as $R \propto
E_e^{-1/2}$.

Option (iii) - a time-variable acceleration spectrum - is a distinct
possibility in particular for accelerated electrons. Higher pulsar
spin-down luminosity in the past will have been associated with higher
$B$ fields and a lower cutoff energy, governed by the relation between
acceleration and radiative cooling time scales.  In either case (i),
(ii) or (iii), the new H.E.S.S.\ results therefore provide evidence of
an electronic origin of the VHE $\gamma$-ray emission, and require
that characteristic cooling time scales are, or at some earlier time
were, shorter than the age of the nebula.

\subsection{Conclusion}
We have presented detailed morphological and spectral studies of the
VHE $\gamma$-ray source HESS\,J1825--137 that has been originally
detected in the survey of the inner Galaxy, conducted by H.E.S.S.\ in
2004. The $\gamma$-ray spectrum of the source has been measured over
more than two decades between $\sim$270~GeV and $\sim$35~TeV. The
energy spectrum shows indications for a deviation from a pure
power-law. Several spectral shapes have been applied to fit the data
and it seems, that a broken power-law or a power-law with energy
dependent photon index provide a better description than a pure power
law. The large data set has provided the possibility for a spatially
resolved spectral study. A significant softening of the $\gamma$-ray
spectrum away from the position of the energetic pulsar
PSR\,J1826--1334 has been found, providing the first direct evidence
of an energy dependent morphology in VHE $\gamma$-rays. The studies
performed here significantly strengthen the case that the VHE
$\gamma$-ray emission originates in the wind nebula of
PSR\,J1826--1334.  It is difficult to explain the measured
$\gamma$-ray luminosity in terms of the current spin-down luminosity
of the pulsar.  A like scenario is a significant contribution of
`relic' electrons released in the early history of the pulsar, when
the spin-down luminosity is higher.  The variation of index with
distance from the pulsar is attributed both to IC and synchrotron
cooling of the continuously accelerated electrons.
\begin{acknowledgements}
  The support of the Namibian authorities and of the University of
  Namibia in facilitating the construction and operation of H.E.S.S.\
  is gratefully acknowledged, as is the support by the German Ministry
  for Education and Research (BMBF), the Max Planck Society, the
  French Ministry for Research, the CNRS-IN2P3 and the Astroparticle
  Interdisciplinary Programme of the CNRS, the U.K. Particle Physics
  and Astronomy Research Council (PPARC), the IPNP of the Charles
  University, the South African Department of Science and Technology
  and National Research Foundation, and by the University of
  Namibia. We appreciate the excellent work of the technical support
  staff in Berlin, Durham, Hamburg, Heidelberg, Palaiseau, Paris,
  Saclay, and in Namibia in the construction and operation of the
  equipment.
\end{acknowledgements}

\bibliographystyle{aa}

\end{document}